\documentclass[11pt,twoside]{article}
\usepackage{./asp2014}

\aspSuppressVolSlug

\resetcounters

\begin{document}

\title{$\gamma$-ray binaries : a bridge between Be stars and high energy astrophysics}
\author{Astrid Lamberts 
\affil{Center for Gravitation, Cosmology and Astrophysics, University of Wisconsin-Milwaukee,  USA; {lambera@uwm.edu}}}

\paperauthor{Astrid Lamberts}{lambera@uwm.edu}{}{Center for Gravitation, Cosmology and Astrophysics}{University of Wisconsin-Milwaukee}{Milwaukee}{Wisconsin}{53201}{USA}



 \newcommand\gb{$\gamma$-ray binary }
 \newcommand\gbs{$\gamma$-ray binaries }








\begin{abstract}

Advances in X-ray and gamma-ray astronomy have opened a new window on our universe and revealed a wide variety of binaries composed of a compact object and a Be star. In Be X-ray binaries, a neutron star accretes the Be disk and truncates it through tidal interactions. Such systems have important X-ray outbursts, some related to the disk structure. In other systems, strong gamma ray emission is observed. In $\gamma$-ray binaries, the neutron star is not accreting but driving a highly relativistic wind. The wind collision region presents similarities to colliding wind binaries composed of massive stars. The high energy emission is coming from particles being accelerated at the relativistic shock.
I will review the physics of X-ray and gamma-ray binaries, focusing particularly on the recent developments on gamma-ray binaries. I will describe physical mechanisms such as relativistic hydrodynamics, tidal forces and non thermal emission.  I will highlight how high energy astrophysics can shed a new light on Be star physics and vice-versa \footnote{A video of the talk can be found at http://activebstars.iag.usp.br/index.php/talk-conference-recordings/bestars-2014-in-london-ontario/session-6/video/astrid-lamberts-interacting-binaries-be-stars-and-high-energy-astrophysics}.


\end{abstract}
\vspace{-1. cm}
\section{Introduction}

Be stars are a subgroup of B-type stars which spectra display one or more Balmer emission lines at some point of their life \citep{1987pbes.coll....3C}.  The emission arises from a dense circumstellar disk.  Be disks are well described by the viscous decretion disk  model \citep{1991MNRAS.250..432L}.  They vary on timescales of a few years \citep{2011MNRAS.413.1600R}. They undergo growth and decay phases \citep{2008MNRAS.386.1922J} and can also present warping due to one-armed oscillations \citep{1991PASJ...43...75O,1992A&A...265L..45P,1994A&A...288..558T}. Complete  reviews on Be stars and disks can be found in \citep{2003PASP..115.1153P,2013A&ARv..21...69R}.

The majority of massive stars  are located in binary systems \citep{2007ApJ...670..747K}. Binary interactions are important during the main sequence, as colliding stellar winds, but also during later stages, when one of the stars has become a compact object.  This review focuses on systems where the compact object is a neutron star, which covers the vast majority of binary systems with Be stars. There is currently only one binary potentially composed of  a black hole \citep{2012MNRAS.421.1103C}. 

\begin{figure}[h]
  \centering
  \includegraphics[width = .75\textwidth ]{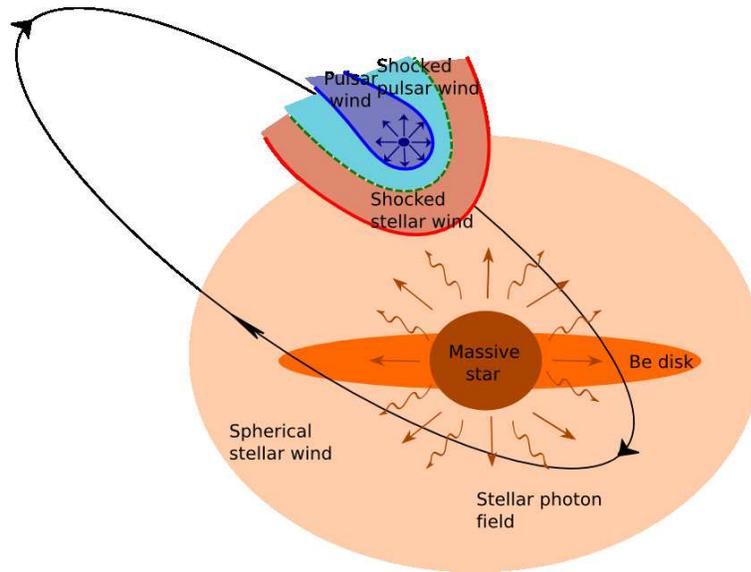}
  \caption{Schematic view of a $\gamma$-ray binary (not to scale). The collision between the pulsar wind and stellar wind creates two shocks. A  contact discontinuity separates both shocked winds.  The photon field of the massive star (and its Be disk if present) provides the seed photons for the inverse Compton emission. The orbit of the pulsar is likely inclined with respect to the Be disk.  }
  \label{fig:gamma_binary}
\end{figure}

After its formation, the neutron star is rotating very fast and drives a highly relativistic wind. The collision between the pulsar wind and wind from the massive companion creates a shocked region where particles are accelerated, yielding $\gamma$-ray emission. A schematic view of the so-called $\gamma$-ray binaries is given in Fig. \ref{fig:gamma_binary}. Five of such systems has been discovered so far, three of them hosting a Be star. The study of such systems is very recent and became possible thanks to the \textit{Fermi} satellite as well as ground-based Cerenkov telescopes such as H.E.S.S., MAGIC or VERITAS. \citet{2013A&ARv..21...64D} describes our current understanding of them. 

As the pulsar slows down, the power of the wind decreases and eventually the neutron star starts accreting its massive companion. The emission is now dominated by X-rays, and such systems are called high-mass X-ray binaries.  More than a hundred of these systems have been detected \citep{2006A&A...455.1165L}, the majority of them is likely hosting  a Be star, dominating the optical emission.  A recent review can be found in  \citet{2011Ap&SS.332....1R}.

High-mass X-ray binaries and gamma-ray binaries are two different phases of the evolution of a massive binary. Good progress in the field of Be/X-ray binaries has been made since the dynamics of the Be disk are better understood and well modeled in simulations of Be/X-ray binaries.  This has extended the interest of Be/X-ray binaries from the high energy community to the larger scope of stellar physics. The understanding of $\gamma$-ray binaries, which is still incomplete,   is bound to follow the same track if we want to fully understand all the observed properties. It will bring valuable information on high energy processes as well as fundamental stellar and disk properties.

This review, based on a talk at the ``Bright emissaries'' conference held in London, Ontario on Aug 10-13 2014, highlights the relevance of the Be phenomenon to high energy astrophysics and vice-versa. The review focuses on the aspects of stellar and disk physics in $\gamma$-ray binaries, and I refer the reader to \citet{2013A&ARv..21...64D} for details on aspects such as particle acceleration or pulsar wind physics.  I will briefly recall the importance of the Be phenomenon in the understanding of Be/X-ray binaries (\S\ref{sec:XB}). Then I will describe the wide variety of observations of $\gamma$-ray binaries (\S 3) and how theoretical models attempt to explain all the observed variability (\S 4). Finally I will recall the importance for improved modeling of stellar physics in the study of $\gamma$-ray binaries (\S 5) and conclude (\S 6).


\section { Learning from Be/Xray binaries}\label{sec:XB}

 On top of quiescent low X-ray luminosity (L$_X\simeq 10^{34} $ erg s$^{-1}$), two types of bursts are observed in certain systems. Type I bursts are quasiperiodic, with $L_X\simeq 10^{36-37}$ erg s$^{-1}$ and are short compared to the orbital period. Type II bursts are more luminous ($L_X\ge 10^{37}$ erg s$^{-1}$),  aperiodic and cover a significant fraction of the orbital period \citep{1986ApJ...308..669S,1998A&A...338..505N}.


The presence of the compact object strongly affects the Be disk, which is perturbed by tidal torques when the viscous timescale is larger than the orbital period. The perturbation leads to disk truncation at a resonance radius  \citep{2002MNRAS.337..967O}. Be disks in Be/X-ray binaries are smaller but denser than disks around single stars \citep{2001A&A...367..884Z}.  In highly eccentric binaries, the truncation radius can be large enough to allow capture of the disk material during the periastron passage of the compact object, resulting in a Type I burst \citep{2001A&A...377..161O}.

The exact origin of type II burst is not clearly established, although the timescales between and during bursts indicate a relation to the dynamics of the disk (rather than the orbital period of the system). The increased accretion occurring during type II burst may result from a warp in the disk, caused by a misaligned orbit of the neutron star \citep{2011MNRAS.416.2827M,2014ApJ...790L..34M}. Detailed studies of H$\alpha$ and HeI emission lines in the disk correlate the presence of asymmetries in the  disk with bursts \citep{2011PASJ...63L..25M}. 

The combination of numerical simulations showing the long term evolution of the disk, theoretical progress on tidal interactions and viscous disks and long-term monitoring campaigns in X-rays and optical spectroscopy led to the current understanding of Be/X-ray binaries. Be/X-ray binaries are not only  a window on the physics of neutron stars and associated accretion and high energy emission.  They are also a highly valuable testbed for Be disks and their evolution.  Careful study of the X-ray bursts provides indications on the density and dynamics of the disks. The relevant timescales provide insights on the viscosity in the disk \citep{2012ApJ...744L..15C} and the central star. The large number of discovered systems allows statistical studies of orbital and spin properties, which gives valuable information of the progenitor system and supernova kicks \citep{2002ApJ...574..364P,2009MNRAS.397.1563M}.



\section {$\gamma$-ray binaries : puzzling observations}
\begin{figure}[h]
  \centering
  \includegraphics[width = .45\textwidth ]{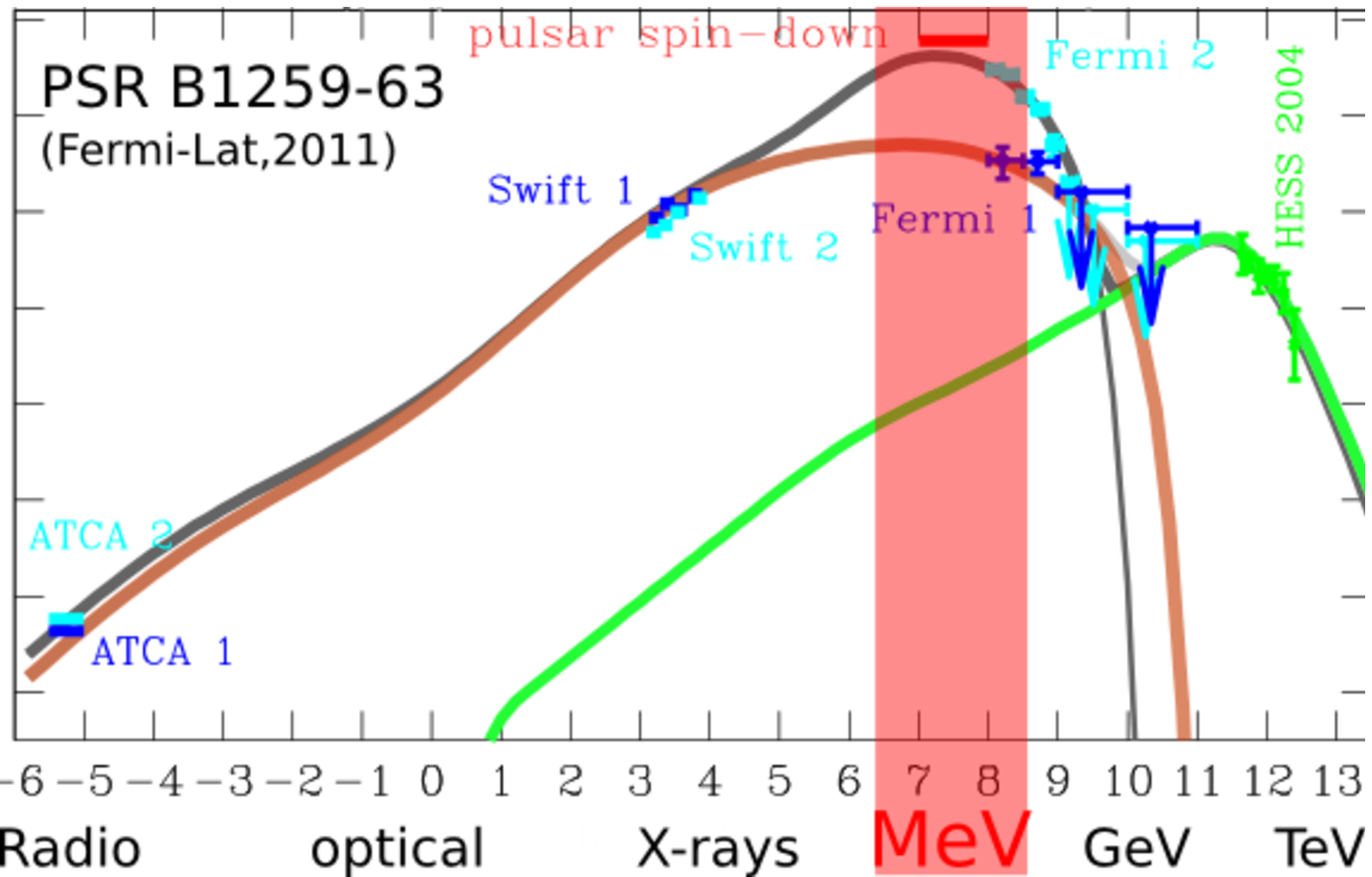}
\includegraphics[width=.4\textwidth]{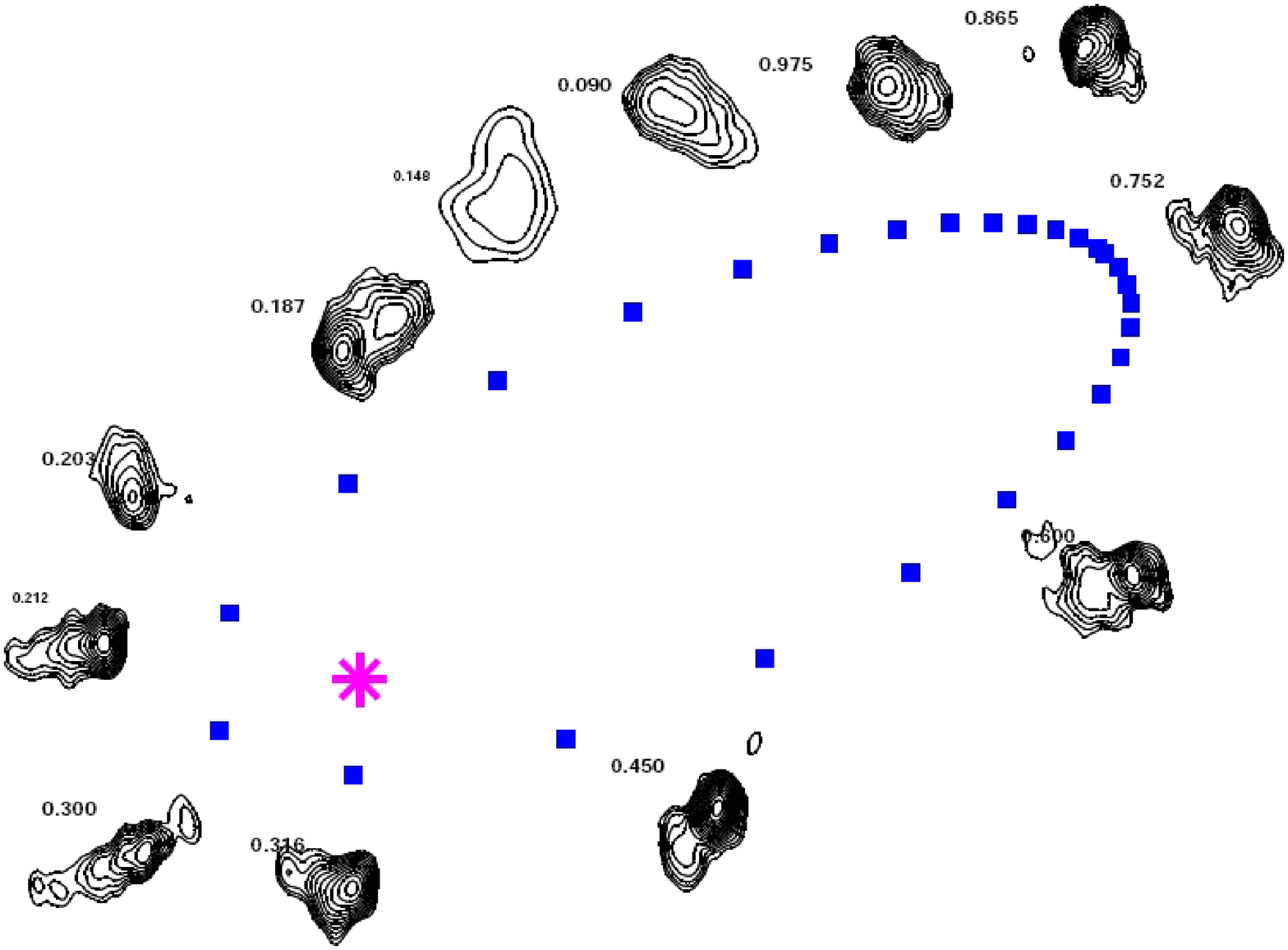}
  \caption{Left : Spectral energy distribution of PSR B1259-63 (Adapted from \citet{2011ApJ...736L..11A} with permission of the authors). Right : radio map of LSI+61$^o$303 (taken from \citet{2006smqw.confE..52D} with permission of the authors).}
  \label{fig:obs}
\end{figure}

The left panel of Fig. \ref{fig:obs} shows the spectral energy distribution of PSR B1259-63, the only $\gamma$-ray binary with a confirmed pulsar. It shows continuous emission from radio up to tens of TeV. The peak of the emission occurs in the MeV range, leading to the denomination of $\gamma$-ray binary. The optical emission is dominated by the massive companion. In all cases, the companions are highly luminous massive stars (from B0Ve to 06V). The non-thermal emission is synchrotron (up to GeV band) and inverse Compton emission (up to TeV) from particles accelerated at the shocks between the winds.  In systems with small orbits, the synchrotron radio emission is resolved at all phases and traces the position of the shocked pulsar wind as it evolves around the companion star. The example of LSI +61$^o$303 is shown on  the right panel of Fig. \ref{fig:obs}. 

Table 1 summarizes the observed variability and main parameters of  in all detected $\gamma$-ray binaries. \gbs strike by their complex variability pattern. Despite of the tremendous progress, current models are still unable to explain all variability in a consistent manner.  Understanding the orbital and superorbital modulations provides a unique chance to probe both the pulsar wind and the stellar environment. 

\begin{table}\label{tab:systems}
\caption{Summary of the main characteristics of the currently known $\gamma$-ray binaries. From left to right : confirmed identification of compact object, spectral type of the companion star, orbital period (in days), eccentricity, and presence of variability in various wavebands, related to the orbital phase (O) and/or another timescale (V), likely related to the Be disk. The presence of parentheses indicate a strong indication without firm confirmation. For references, see \citet{2013A&ARv..21...64D}. The observation of a magnetar-like flare is a strong indication for the presence of a neutron star in LSI + 61$^o$ 303 \citep{2008ATel.1715....1D}. }

\begin{center}
\begin{tabular}{c  c c c c c c c c c }
\hline
\noalign{\smallskip}
                    \footnotesize{system}     & \footnotesize{ pulsar} &\footnotesize{star}  &  \footnotesize{ P$_{orb}$}&\footnotesize{ e}     &\footnotesize{ radio }& \footnotesize{$H_{\alpha}$} & \footnotesize{ X }& \footnotesize{ GeV }& \footnotesize{ TeV} \\ 
\noalign{\smallskip}
\hline
\noalign{\smallskip}
\footnotesize{PSR B1259-63}  & X & O9.5Ve & 1237  & 0.87 & O&  O&O & O & O\\
\footnotesize{LSI + 61$^o$ 303} &(X) &B0Ve & 26.5 & 0.54  &  O/V&O/V & O/V & O/V & O/(V)\\
\footnotesize{ LS 5039} &   & O6.5V& 3.9 & 0.35 & O &  & O & O &O\\
\footnotesize{HESS J0632+057} &  & B0Ve& 321 & 0.83  & O &O/V &O && O \\
\footnotesize{1FGL J1018.6-5856} & & O6V&16.6 & ? & O & &O &O&O \\
\noalign{\smallskip}

\hline
\end{tabular}
\end{center}
\end{table}

\begin{figure}[h]
  \centering
  \includegraphics[width = .45\textwidth ]{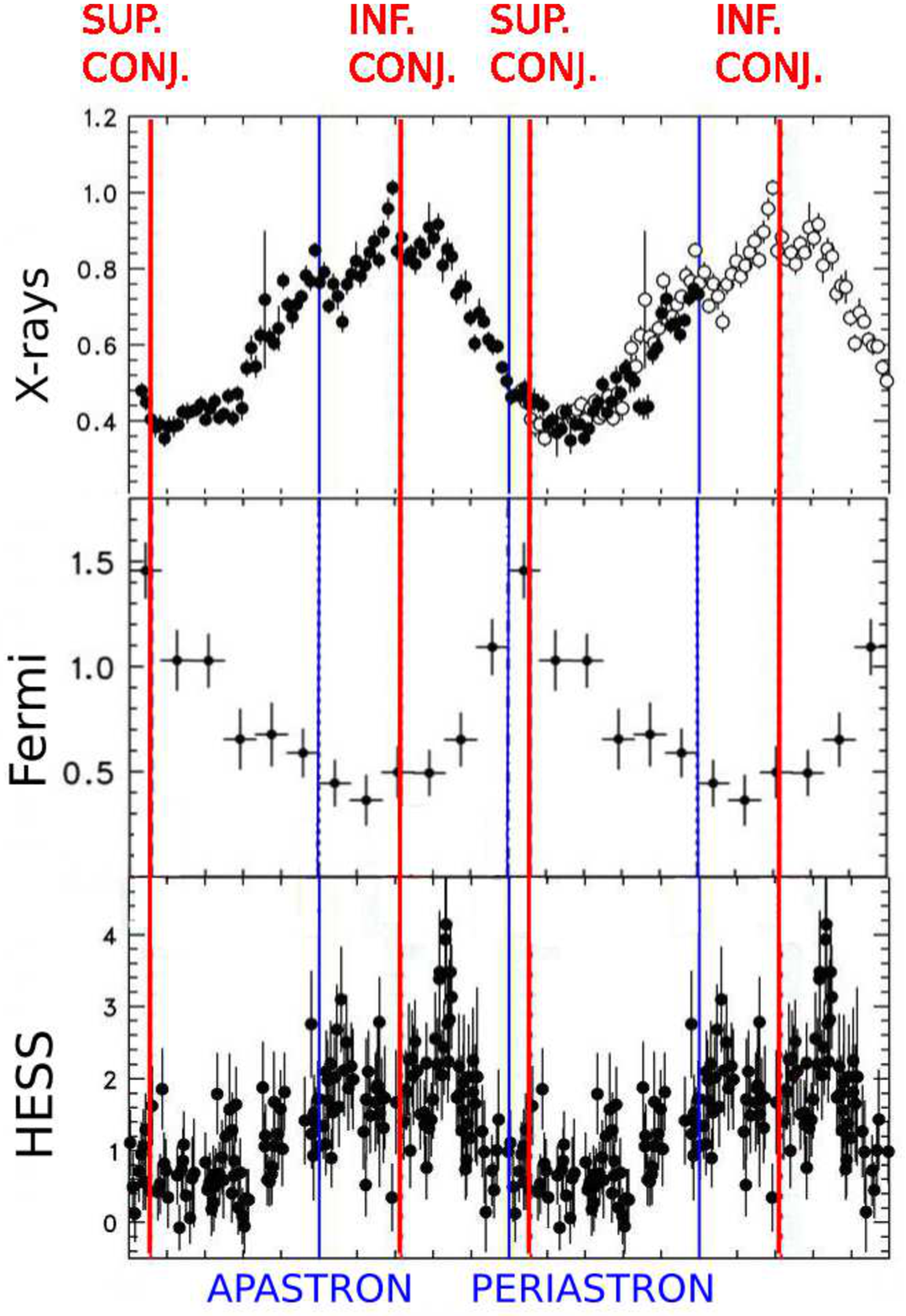}
  \caption{X-ray, GeV and TeV modulation for LS 5039. Adapted from \citet{2009ApJ...697L...1K,2009ApJ...706L..56A,2006A&A...460..743A} with permission of the authors.}
  \label{fig:curve_LS}
\end{figure}

LS 5039 is probably the best understood \gb thanks to its short orbital period and the absence of the Be disk. Fig. \ref{fig:curve_LS} shows its lightcurves in  X-rays, GeV and TeV.  Radio, X-rays and the TeV modulations display the same behavior, peaking around inferior conjunction and showing the least emission at superior conjunction, when the compact object is behind the massive star with respect to the observer. \textit{Fermi} observations on the other hand peak around periastron and display the lowest luminosity close to apastron. The peculiar variability of the GeV emission, confirmed in other systems, is one of the challenges of theoretical models. Peak/lows at conjunctions rather than periastron and apastron indicate variability is related to geometrical effects rather than intrinsic variability due to changes in the shocked region.  

Due the large eccentricity and orbital period, PSR B1259-63  shows non-thermal emission mainly around periastron. TeV, X-rays, radio and $H_{\alpha}$ display increased emission about 20 days before periastron and another peak about 20 days after  \citep{2014MNRAS.439..432C}. The timing corresponds to the pulsar passing trough the inclined Be disk \citep{1997ApJ...477..439T,2007MNRAS.380..320K}, although the exact emission pattern is not fully understood.  The rise before periastron, and slow decline afterwards in the equivalent width (EW) of Br$\gamma$ and H$\alpha$ indicate a growth and decay of the Be disk \citep{2006ApJ...651L..53G}.  More observations are necessary to determine timescales and line profile variability. About 60 days after periastron, there is strong flaring in the GeV range \citep{2011ApJ...736L..11A}, while there is no increased emission at other wavebands. The presence of the flare was confirmed after the last periastron passage in 2014 \citep {2014ATel.6216....1T}. It shows  energy very close to the pulsar's spindown power and remains  unexplained  \citep{2013A&A...557A.127D,2012ApJ...752L..17K,2014JHEAp...3...18S}. 



\begin{figure}[h]
  \centering
  \includegraphics[width = .4\textwidth ]{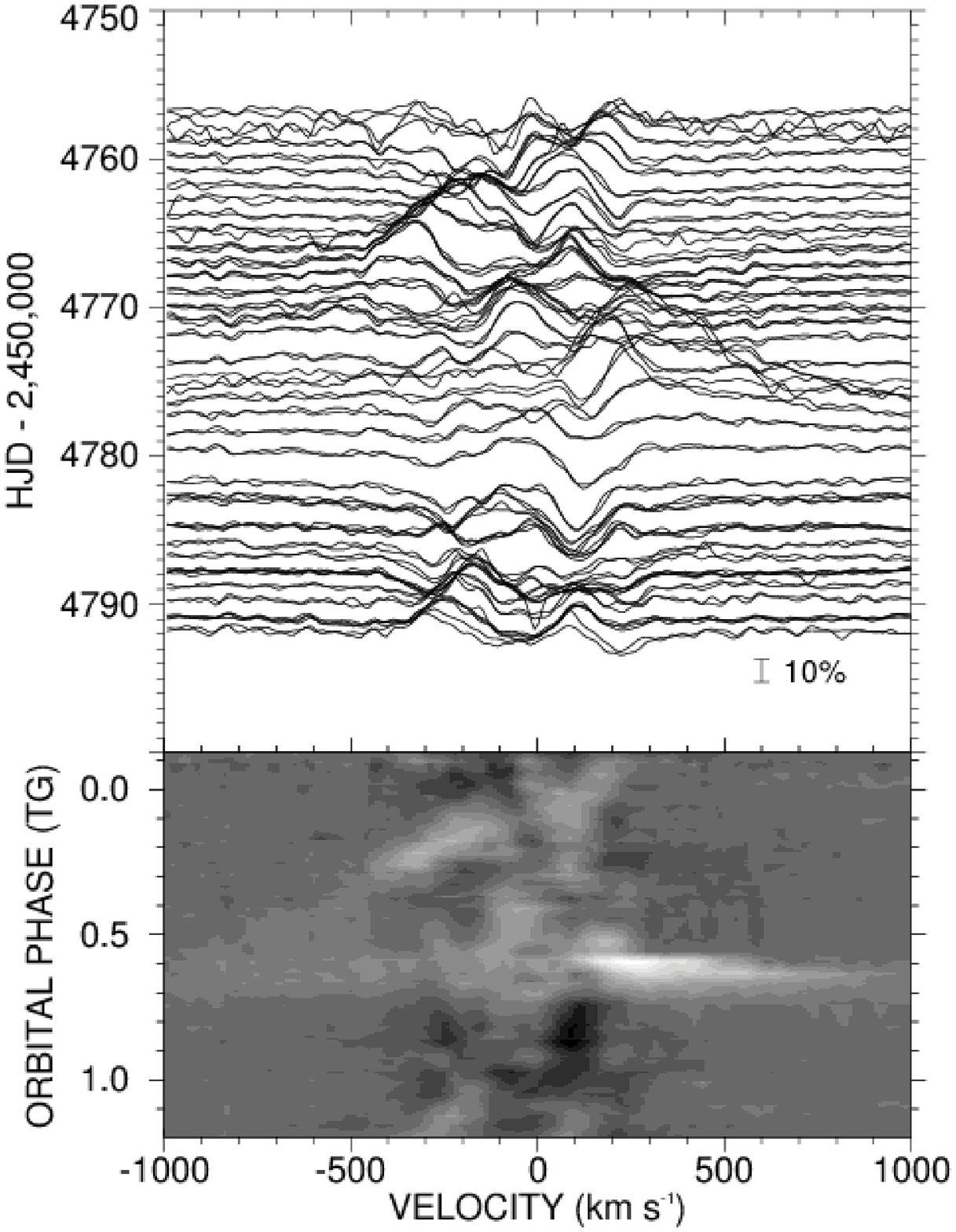}
  \includegraphics[width = .45\textwidth ]{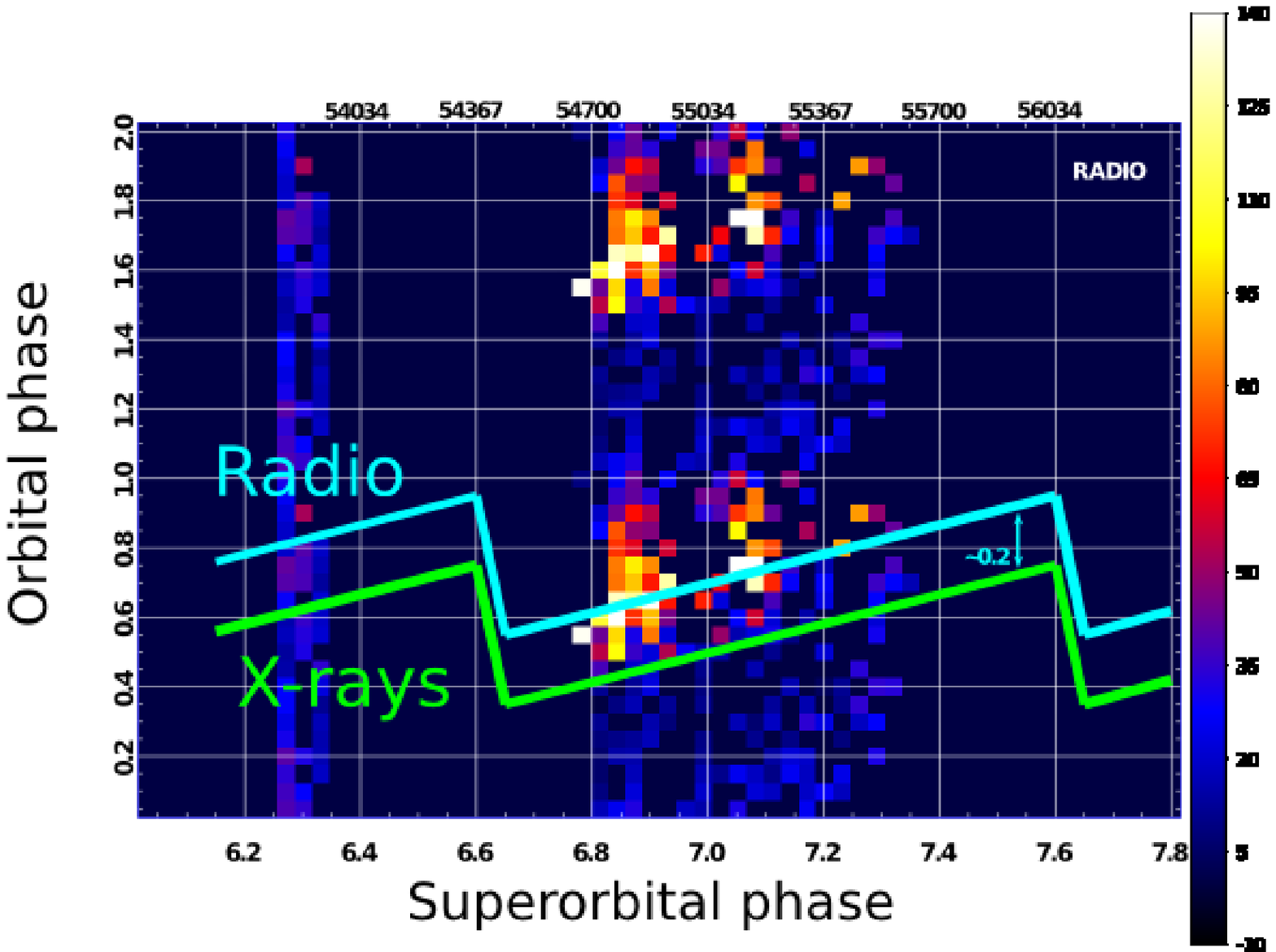}
  \caption{Left : Emission residuals for the H$\alpha$ line over a full orbit \citep{2010ApJ...724..379M}. Right : Superorbital variation of the radio and X-ray flux in LSI+ 61$^o$ 303 \citep{2012ApJ...747L..29C}. Reproduced with permission of the authors.}
  \label{fig:LSI}
\end{figure}

LSI+61$^o$303 is another binary with a Be star, which a much shorter orbit. The orbital variability of the H$\alpha$ line profile is shown on the  left panel of Fig. \ref{fig:LSI} \citep{2010ApJ...724..379M}. It shows a S-pattern over the second half of the orbit, indicating the presence of a one armed spiral density wave in the disk \citep{2003PASP..115.1153P}. The red shoulder before apastron is not originating from the Be disk itself and its origin is not explained.
Following the results on X-ray binaries, we can assume that the pulsar does not directly interact with the disk in this system but probably strongly truncates it. It may be responsible for the observed superorbital variability as well, as both the tidal perturbation and the hydrodynamic interaction can lead to warping of the disk.  The systems has a superorbital period of 1667 days \citep{2002ApJ...575..427G,2013arXiv1307.6384T}, where the peak of the emission shifts \citep{2012ApJ...747L..29C} by about half an orbit.  The right panel of Fig. \ref{fig:LSI} shows the peak of the radio and X-ray emission as a function of the orbital and superorbital phase.  This implies only observations with the same orbital and superorbital period  can be safely combined for comparison with theoretical predictions.

The third $\gamma$-ray binary with a Be star is HESS J0632+057 \citep{2007A&A...469L...1A}, which long term variability is still unclear. The variability of the EW(H$_{\alpha}$) with a period of $\simeq$ 60 days \citep{2010ApJ...724..306A}, suggest the presence of a one-armed oscillation in the disk. So far, this system strikes by the absence of GeV emission, even after deep searches \citep{2013MNRAS.436..740C}. This again, suggests a different origin for the GeV emission and emission at all other wavebands.

In the next section, I will describe our current understanding of $\gamma$-ray binaries and the theoretical challenges, some of them being directly related to the Be star.


\section { Theory : success and challenges}
The colliding wind paradigm, proposed by  \citet{1995MNRAS.275..381M,1994ApJ...433L..37T}  and brought back to the front scene by \citet{2006A&A...456..801D} is now well established. The collision between the dense stellar outflow and the highly relativistic, tenuous pulsar wind results in a shocked structure where leptons are accelerated up to very high energies.  The particles then cool adiabatically and radiatively as they are advected away from the shocks. The very high energy emission results from inverse Compton scattering of the stellar and disk photons \citep{1999APh....10...31K}.  The lower energy emission, peaking in the MeV range and extending down to radio wavelengths is synchrotron emission.

 Using this model as a baseline, one can analytically determine the position of the shocked region  between both winds (see Fig. \ref{fig:gamma_binary}) and estimate the values of hydrodynamical quantities such as density, velocity \citep{1996ApJ...469..729C} and magnetic field \citep{1984ApJ...283..694K}.  Based on such one-zone, or one-dimensional models, one can then compute the resulting emission. Various models take into account  adiabatic losses in the shocked  wind \citep{2007MNRAS.380..320K,2009ApJ...697..592T,2011A&A...527A...9Z},  the cascade of non-thermal particles \citep{2010A&A...519A..81C,2006MNRAS.368..579B}, emission from the unshocked pulsar wind \citep{2011ApJ...742...98K,2008A&A...488...37C} and very high energy gamma-ray absorption due to pair production \citep{2006A&A...451....9D,2009ApJ...693.1462S}.   Variations of the standoff point, due to the eccentricity of the binary,  combined with  line-of-sight effects naturally yield orbital modulations. 

Complementary studies have focused on the geometry and dynamics of the interaction region, using multidimensional hydrodynamical simulations. Such simulation reveal the close proximity to colliding winds of massive stars. Colliding stellar winds have been studied for decades (see e.g. \citet{ 2009MNRAS.396.1743P})  and their modeling is computationally less demanding than $\gamma$-ray binaries.  Simulations highlight the presence of instabilities in the colliding wind region \citep{2011MNRAS.418.2618L}, leading to mixing between both winds, which increases Coulomb losses of the high-energy particles \citep{2010MNRAS.403.1873Z}. \citet{2011PASJ...63..893O} studied the interaction between the Be disk and the pulsar using 3D smoothed particle hydrodynamics (SPH) simulations. They find that, similarly to X-ray binaries, the Be disk has to be denser than disks in isolated stars.



The pulsar wind is highly relativistic, with a Lorentz factor $\Gamma=(1-v^2)^{-1/2}\simeq 10^{5-7}$. Relativistic hydrodynamics differs from classical hydrodynamics by the presence of the Lorentz factor in the conservation equations. This yields a more complex coupling between density, energy and momentum in all spatial directions, which can affect the shocked structure. The equation of state is also more elaborate \citep{2007MNRAS.378.1118M}. Properly modeling the shocked pulsar wind is necessary to determine its Lorentz factor, which is critical to determine Doppler boosting.  The emission of a relativistic is enhanced when the flow is aligned with the line of sight.  This effect is likely the cause of the orbital variations of the TeV lightcurve in LS 5039 \citep{2010A&A...516A..18D}.

Taking into account the relativistic nature of the pulsar wind in a high resolution, large scale numerical simulation remains a computational challenge. Numerical relativistic hydrodynamics is still an active field of research and current simulations are limited to a Lorentz factor $\Gamma\simeq 10-30$. Simulations are computationally demanding as the dynamical  timescale for the pulsar wind (which sets the size of the time-step) is about a hundred times smaller than the timescale for the stellar wind, not to mention the timescales for the Be disk and orbital period. 

\begin{figure}[h]
  \centering
 \includegraphics[trim= 0.cm 0cm 0cm 0cm, clip,width = .45\textwidth ]{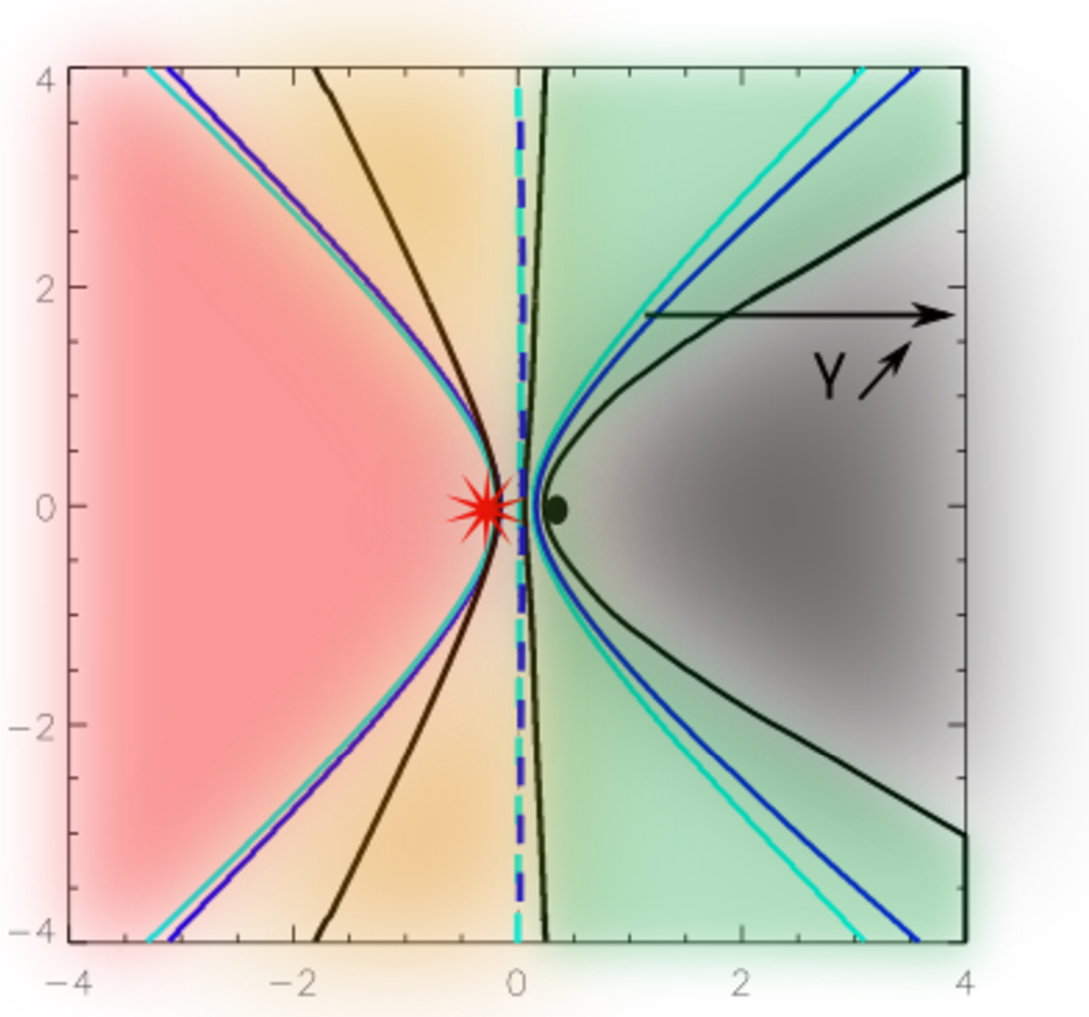} 
 \includegraphics[trim= 0cm 0cm 0cm 0cm, clip,width = .45\textwidth ]{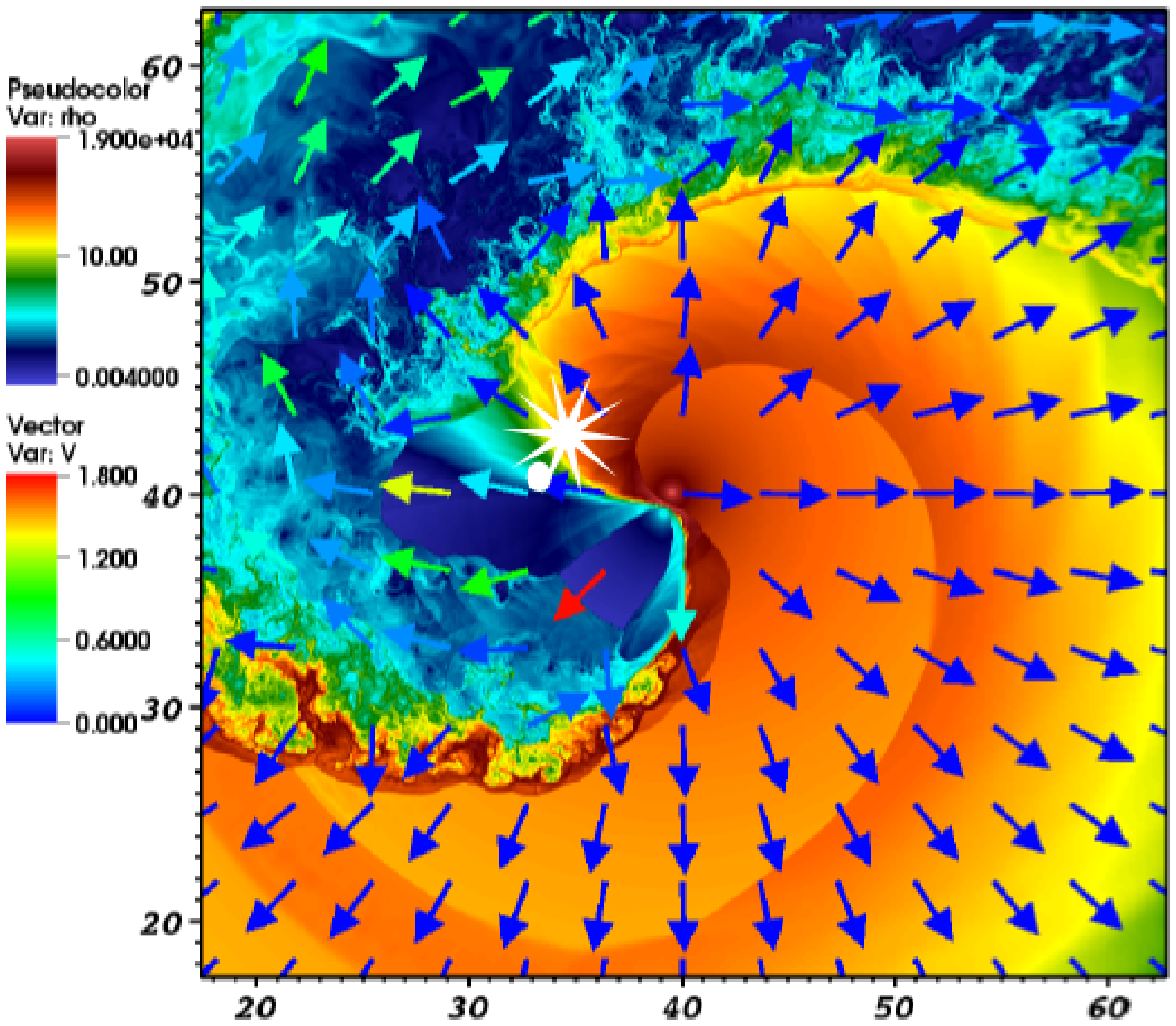} 

  \caption{Left : Position of the stellar wind shock, contact discontinuity and pulsar wind shock (from left to right) for increasing values of $\Gamma$ : 1 (pale blue), 3 (dark blue) and 7 (black). Adapted from \citet{2013A&A...560A..79L}. Right : large scale structure of LS 5039. The pulsar wind ($\Gamma=2$) is the blue region, the stellar wind the orange one. Taken from \citet{2012A&A...544A..59B}. In both cases, the lengthscale is set by the binary separation. Reproduced with permission of the authors.}
  \label{fig:simu}
\end{figure}

Performing relativistic simulations, where the shocks are set as numerical boundaries, \citet{2008MNRAS.387...63B} find that the shocked pulsar wind can re-accelerate up to $\Gamma\simeq 100$ at a distance of $~50$ times the binary separation.  In smaller scale follow-up simulations, they find little impact of the magnetic field and the anisotropy of the pulsar wind \citep{2012MNRAS.419.3426B}. The left panel of Fig. \ref{fig:simu} shows that relativistic coupling between spatial directions leads to a more confined pulsar wind \citep{2013A&A...560A..79L}. The right panel shows that, at larger scales the  spiral structure \citep{2012A&A...544A..59B} is very similar to the non-relativistic case \citep{2012A&A...546A..60L}, although this has to be confirmed at higher Lorentz factors.

In spite of these theoretical efforts, there is currently no fully consistent model which is able to explain all variations at all wavelengths. First,  the  peculiar behavior of the GeV lightcurves, with respect to other wavebands suggests the existence of two distinct populations of energetic particles \citep{2013A&A...551A..17Z}, which origin remains unconfirmed.  Similarly, there is no satisfactory explanation for the  GeV flare observed in PSR B1259-63 \citep{2013A&A...557A.127D,2012ApJ...752L..17K}.  Finally, the confirmed superorbits in binaries with Be stars pinpoint interactions between the disk and the compact object \citep{2011PASJ...63..893O} but studies are too scarce to allow quantitative understanding. 

The full benefit of the wealth of currently available data can only be obtained when the many aspects of theoretical modeling come together.  Progress will hopefully be made by bringing together simulations, fully capturing the structure of $\gamma$-ray binaries, and the models for non-thermal emission. This should allow to identify the relevant emission regions. Similarly, including more elaborate models for the pulsar wind, its magnetic field, geometry and composition will lead to a better model for the injection of non-thermal particles.  Finally, important progress can be made by using an more updated vision for the companion star : its spherical wind, photon field and definitely the Be disk itself. In the next section, I will detail the need for improved modeling of stellar and disk physics in $\gamma$-ray binaries but also highlight how such objects can provide important clues for stellar physics.




\section{Importance of the stellar companion}
During their often eccentric and inclined (with respect to the Be disk) orbit, pulsars sample a variety of stellar environments.  In LS 5039 the pulsar is located in different areas of the wind acceleration region.  In PSR B1259-63 (and HESS J0632+057 ?) , the pulsar goes through an increasingly dense, but slower,  stellar wind as it approaches the star and then eventually interacts with the equatorial disk, twice. In all systems, the variations of the photon field impact the inverse Compton emission.

The momentum flux ratio of the winds is the main criterion for the  structure of the colliding wind region.  Uncertainties on the momentum flux ratio relate to the uncertainties on the stellar mass loss rate and the pulsar's spindown power. Observational constrains provide sets of possible values \citep{2007A&A...473..545B,2011MNRAS.411.1293S}.  The non-detection of thermal X-rays, traditional signature of colliding wind regions puts upper limits on both values \citep{2011A&A...527A...9Z}. Similarly, the lack of absorption of the non-thermal X-rays in LS 5039 at superior conjunction puts an upper limit of $\dot{M}=1.5\times 10^7 M_{\odot}$ yr$^{-1}$, combined with the presence of an extended emitting region. Clumps in the winds affect the derived mass loss rate and increases the X-ray variability \citep{2007A&A...473..545B,2010MNRAS.403.1873Z}.

The characteristics of the Be disk are essentially unknown.  Still, the presence of the disk is a crucial aspect of $\gamma$-ray binaries.  In PSR B1259-63, the pulsar likely passes through the disk at periastron. A first attempt at modeling such interaction \citep{2011PASJ...63..893O} highlights that most of the destruction of the disk is related to the pulsar wind, rather than a purely tidally interaction as is found in X-ray binaries.  Only dense disks undergo limited destruction. Based on the same simulations,  \citet{2012ApJ...750...70T} reproduce the double-peaked X-ray light curves but fail to reproduce the TeV lightcurves.  Higher resolution simulations, may be necessary to better resolve the shocked structure around the apex and obtain more satisfactory lightcurves. 

The long term fate of the disk is unexplored so far : does is ``refill''  and recover an unperturbed state while the pulsar is at apastron? If not, how much does the disk shrink and evolve between passages?  These questions may remain unresolved for a while as long-term 3D simulations of $\gamma$-ray binaries are limited by the currently available computer power.

In LSI+61$^o$ 303 the pulsar maybe influences the disk during the whole orbital period, probably with some variable effects due to the eccentricity of the system, as is suggested by the variable H$\alpha$ profiles. Whether the pulsar is directly responsible for the observed stable superorbital modulation is not clear. While single Be stars display variability on similar timescales due to one armed oscillations and/or warping, the variability is not always cyclic \citep{1997A&A...318..548O}. Why is the presence of the cpulsar companion stabilizing the variability of the disk and setting that particular timescale?  Fully modeling the disk, as well as its interactions with the central star, \textit{\`{a} la} \citet{2012ApJ...756..156H}, while it is feeling an outside tidal field as well as hydrodynamical interaction might answer such questions.

The disk is also an important source of infrared photons, which can increase the inverse Compton emission  \citep{2012MNRAS.426.3135V,2011MNRAS.412.1721V}, which may be one of the necessary ingredients to explain the GeV flare in PSR B1259-63.

Finally, 3 out of 5 known systems are hosting a Be star. While this may be just a result of the small sample, which has little statistical significance,  it may open new leads on the fundamental question of the origin of the Be phenomenon itself.
\section{Some conclusions, more perspectives }
Since the confirmation of TeV emission in PSR B1259-63 \citep{2005A&A...442....1A}, the field of $\gamma$-ray binaries has made a tremendous progress. The \textit{Fermi} satellite has revealed the peculiar nature of the GeV emission. Long-term multi-wavelength monitoring shows orbital variability at all wavelengths. Systems with a Be star display additional modulations, observed at most wavebands as well. The combination of data has established the current paradigm : the colliding wind region allows for particle acceleration, which results in high energy emission through inverse Compton and synchrotron emission.  Some modulations can be explained by geometrical effects and changes in the wind collision structure as the pulsar sample a changing environment of the massive star. Similarly superorbital modulations are related to the Be disk.

Still, many questions remain unanswered. The structure and composition of the  pulsar wind, its magnetic field and the injection of non-thermal particles is not well constrained.  The exact location of the emitting regions is not clear, although there is increasing evidence for extended emission and/or multiple emitting regions.  While  the interaction between the Be disk and compact object are responsible for the emission around periastron in PSR B1259-63 and the superorbital variations in LSI+61$^o$303, the long term evolution of the disk and origin of the superorbit are unknown.

Similarly to   X-ray binaries, further progress in the understanding of $\gamma$-ray binaries will be facilitated through dialogue between the different fields of research.  While plasma physics can likely answer the questions related to the pulsar wind physics, stellar physics is necessary to understand the structure of the colliding wind region and its long term evolution.  This will lead to progress in high-energy astrophysics but also allow to fully exploit all the currently available data on the Be phenomenon.

\section*{Acknowledgments}


\bibliographystyle{asp2014}
\bibliography{biblio_be}

\begin{thebibliography}{}
\expandafter\ifx\csname natexlab\endcsname\relax\def\natexlab#1{#1}\fi
\expandafter\ifx\csname url\endcsname\relax
  \def\url#1{\texttt{#1}}\fi
\expandafter\ifx\csname urlprefix\endcsname\relax\def\urlprefix{URL }\fi
\providecommand{\eprint}[2][]{\url{#2}}

\bibitem[{{Aragona} et~al.(2010){Aragona}, {McSwain}, \& {De
  Becker}}]{2010ApJ...724..306A}
{Aragona}, C., {McSwain}, M.~V., \& {De Becker}, M. 2010, \apj, 724, 306.
  \eprint{1009.2100}

\bibitem[{{Bednarek}(2006)}]{2006MNRAS.368..579B}
{Bednarek}, W. 2006, \mnras, 368, 579. \eprint{astro-ph/0601657}

\bibitem[{{Bogovalov} et~al.(2012){Bogovalov}, {Khangulyan}, {Koldoba},
  {Ustyugova}, \& {Aharonian}}]{2012MNRAS.419.3426B}
{Bogovalov}, S.~V., {Khangulyan}, D., {Koldoba}, A.~V., {Ustyugova}, G.~V., \&
  {Aharonian}, F.~A. 2012, \mnras, 419, 3426. \eprint{1107.4831}

\bibitem[{{Bogovalov} et~al.(2008){Bogovalov}, {Khangulyan}, {Koldoba},
  {Ustyugova}, \& {Aharonian}}]{2008MNRAS.387...63B}
{Bogovalov}, S.~V., {Khangulyan}, D.~V., {Koldoba}, A.~V., {Ustyugova}, G.~V.,
  \& {Aharonian}, F.~A. 2008, \mnras, 387, 63. \eprint{0710.1961}

\bibitem[{{Bosch-Ramon} et~al.(2012){Bosch-Ramon}, {Barkov}, {Khangulyan}, \&
  {Perucho}}]{2012A&A...544A..59B}
{Bosch-Ramon}, V., {Barkov}, M.~V., {Khangulyan}, D., \& {Perucho}, M. 2012,
  \aap, 544, A59. \eprint{1203.5528}

\bibitem[{{Bosch-Ramon} et~al.(2007){Bosch-Ramon}, {Motch}, {Rib{\'o}}, {Lopes
  de Oliveira}, {Janot-Pacheco}, {Negueruela}, {Paredes}, \&
  {Martocchia}}]{2007A&A...473..545B}
{Bosch-Ramon}, V., {Motch}, C., {Rib{\'o}}, M., {Lopes de Oliveira}, R.,
  {Janot-Pacheco}, E., {Negueruela}, I., {Paredes}, J.~M., \& {Martocchia}, A.
  2007, \aap, 473, 545. \eprint{astro-ph/0703499}

\bibitem[{{Caliandro} et~al.(2013){Caliandro}, {Hill}, {Torres}, {Hadasch},
  {Ray}, {Abdo}, {Hessels}, {Ridolfi}, {Possenti}, {Burgay}, {Rea}, {Tam},
  {Dubois}, {Dubus}, {Glanzman}, \& {Jogler}}]{2013MNRAS.436..740C}
{Caliandro}, G.~A., {Hill}, A.~B., {Torres}, D.~F., {Hadasch}, D., {Ray}, P.,
  {Abdo}, A., {Hessels}, J.~W.~T., {Ridolfi}, A., {Possenti}, A., {Burgay}, M.,
  {Rea}, N., {Tam}, P.~H.~T., {Dubois}, R., {Dubus}, G., {Glanzman}, T., \&
  {Jogler}, T. 2013, \mnras, 436, 740. \eprint{1308.5234}

\bibitem[{{Canto} et~al.(1996){Canto}, {Raga}, \&
  {Wilkin}}]{1996ApJ...469..729C}
{Canto}, J., {Raga}, A.~C., \& {Wilkin}, F.~P. 1996, \apj, 469, 729

\bibitem[{{Carciofi} et~al.(2012){Carciofi}, {Bjorkman}, {Otero}, {Okazaki},
  {{\v S}tefl}, {Rivinius}, {Baade}, \& {Haubois}}]{2012ApJ...744L..15C}
{Carciofi}, A.~C., {Bjorkman}, J.~E., {Otero}, S.~A., {Okazaki}, A.~T., {{\v
  S}tefl}, S., {Rivinius}, T., {Baade}, D., \& {Haubois}, X. 2012, \apjl, 744,
  L15. \eprint{1112.0053}

\bibitem[{{Casares} et~al.(2012){Casares}, {Rib{\'o}}, {Ribas}, {Paredes},
  {Vilardell}, \& {Negueruela}}]{2012MNRAS.421.1103C}
{Casares}, J., {Rib{\'o}}, M., {Ribas}, I., {Paredes}, J.~M., {Vilardell}, F.,
  \& {Negueruela}, I. 2012, \mnras, 421, 1103. \eprint{1201.1726}

\bibitem[{{Cerutti} et~al.(2008){Cerutti}, {Dubus}, \&
  {Henri}}]{2008A&A...488...37C}
{Cerutti}, B., {Dubus}, G., \& {Henri}, G. 2008, \aap, 488, 37.
  \eprint{0807.1226}

\bibitem[{{Cerutti} et~al.(2010){Cerutti}, {Malzac}, {Dubus}, \&
  {Henri}}]{2010A&A...519A..81C}
{Cerutti}, B., {Malzac}, J., {Dubus}, G., \& {Henri}, G. 2010, \aap, 519, A81.
  \eprint{1006.2683}

\bibitem[{{Chernyakova} et~al.(2014){Chernyakova}, {Abdo}, {Neronov},
  {McSwain}, {Mold{\'o}n}, {Rib{\'o}}, {Paredes}, {Sushch}, {de Naurois},
  {Schwanke}, {Uchiyama}, {Wood}, {Johnston}, {Chaty}, {Coleiro}, {Malyshev},
  \& {Babyk}}]{2014MNRAS.439..432C}
{Chernyakova}, M., {Abdo}, A.~A., {Neronov}, A., {McSwain}, M.~V.,
  {Mold{\'o}n}, J., {Rib{\'o}}, M., {Paredes}, J.~M., {Sushch}, I., {de
  Naurois}, M., {Schwanke}, U., {Uchiyama}, Y., {Wood}, K., {Johnston}, S.,
  {Chaty}, S., {Coleiro}, A., {Malyshev}, D., \& {Babyk}, I. 2014, \mnras, 439,
  432. \eprint{1401.1386}

\bibitem[{{Chernyakova} et~al.(2012){Chernyakova}, {Neronov}, {Molkov},
  {Malyshev}, {Lutovinov}, \& {Pooley}}]{2012ApJ...747L..29C}
{Chernyakova}, M., {Neronov}, A., {Molkov}, S., {Malyshev}, D., {Lutovinov},
  A., \& {Pooley}, G. 2012, \apjl, 747, L29. \eprint{1203.1944}

\bibitem[{{Collins}(1987)}]{1987pbes.coll....3C}
{Collins}, G.~W., II 1987, in IAU Colloq. 92: Physics of Be Stars, edited by
  A.~{Slettebak}, \& T.~P. {Snow}, 3

\bibitem[{{Dhawan} et~al.(2006){Dhawan}, {Mioduszewski}, \&
  {Rupen}}]{2006smqw.confE..52D}
{Dhawan}, V., {Mioduszewski}, A., \& {Rupen}, M. 2006, in VI Microquasar
  Workshop: Microquasars and Beyond

\bibitem[{{Dubus}(2006{\natexlab{a}})}]{2006A&A...451....9D}
{Dubus}, G. 2006{\natexlab{a}}, \aap, 451, 9. \eprint{astro-ph/0509633}

\bibitem[{{Dubus}(2006{\natexlab{b}})}]{2006A&A...456..801D}
--- 2006{\natexlab{b}}, \aap, 456, 801. \eprint{astro-ph/0605287}

\bibitem[{{Dubus}(2013)}]{2013A&ARv..21...64D}
--- 2013, \aapr, 21, 64. \eprint{1307.7083}

\bibitem[{{Dubus} \& {Cerutti}(2013)}]{2013A&A...557A.127D}
{Dubus}, G., \& {Cerutti}, B. 2013, \aap, 557, A127. \eprint{1308.4531}

\bibitem[{{Dubus} et~al.(2010){Dubus}, {Cerutti}, \&
  {Henri}}]{2010A&A...516A..18D}
{Dubus}, G., {Cerutti}, B., \& {Henri}, G. 2010, \aap, 516, A18.
  \eprint{1004.0511}

\bibitem[{{Dubus} \& {Giebels}(2008)}]{2008ATel.1715....1D}
{Dubus}, G., \& {Giebels}, B. 2008, The Astronomer's Telegram, 1715, 1

\bibitem[{{Fermi-LAT Collaboration}(2009)}]{2009ApJ...706L..56A}
{Fermi-LAT Collaboration} 2009, \apjl, 706, L56. \eprint{0910.5520}

\bibitem[{{Fermi-LAT Collaboration}(2011)}]{2011ApJ...736L..11A}
--- 2011, \apjl, 736, L11. \eprint{1103.4108}

\bibitem[{{Gregory}(2002)}]{2002ApJ...575..427G}
{Gregory}, P.~C. 2002, \apj, 575, 427

\bibitem[{{Grundstrom} \& {Gies}(2006)}]{2006ApJ...651L..53G}
{Grundstrom}, E.~D., \& {Gies}, D.~R. 2006, \apjl, 651, L53.
  \eprint{astro-ph/0609602}

\bibitem[{{Haubois} et~al.(2012){Haubois}, {Carciofi}, {Rivinius}, {Okazaki},
  \& {Bjorkman}}]{2012ApJ...756..156H}
{Haubois}, X., {Carciofi}, A.~C., {Rivinius}, T., {Okazaki}, A.~T., \&
  {Bjorkman}, J.~E. 2012, \apj, 756, 156. \eprint{1207.2612}

\bibitem[{{H.E.S.S. Collaboration}(2005)}]{2005A&A...442....1A}
{H.E.S.S. Collaboration} 2005, \aap, 442, 1. \eprint{astro-ph/0506280}

\bibitem[{{H.E.S.S. Collaboration}(2006)}]{2006A&A...460..743A}
--- 2006, \aap, 460, 743. \eprint{astro-ph/0607192}

\bibitem[{{H.E.S.S. collaboration}(2007)}]{2007A&A...469L...1A}
{H.E.S.S. collaboration} 2007, \aap, 469, L1. \eprint{0704.0171}

\bibitem[{{Jones} et~al.(2008){Jones}, {Sigut}, \&
  {Porter}}]{2008MNRAS.386.1922J}
{Jones}, C.~E., {Sigut}, T.~A.~A., \& {Porter}, J.~M. 2008, \mnras, 386, 1922

\bibitem[{{Kennel} \& {Coroniti}(1984)}]{1984ApJ...283..694K}
{Kennel}, C.~F., \& {Coroniti}, F.~V. 1984, \apj, 283, 694

\bibitem[{{Khangulyan} et~al.(2011){Khangulyan}, {Aharonian}, {Bogovalov}, \&
  {Rib{\'o}}}]{2011ApJ...742...98K}
{Khangulyan}, D., {Aharonian}, F.~A., {Bogovalov}, S.~V., \& {Rib{\'o}}, M.
  2011, \apj, 742, 98. \eprint{1104.0211}

\bibitem[{{Khangulyan} et~al.(2012){Khangulyan}, {Aharonian}, {Bogovalov}, \&
  {Rib{\'o}}}]{2012ApJ...752L..17K}
--- 2012, \apjl, 752, L17. \eprint{1107.4833}

\bibitem[{{Khangulyan} et~al.(2007){Khangulyan}, {Hnatic}, {Aharonian}, \&
  {Bogovalov}}]{2007MNRAS.380..320K}
{Khangulyan}, D., {Hnatic}, S., {Aharonian}, F., \& {Bogovalov}, S. 2007,
  \mnras, 380, 320. \eprint{astro-ph/0605663}

\bibitem[{{Kirk} et~al.(1999){Kirk}, {Ball}, \&
  {Skj{\ae}raasen}}]{1999APh....10...31K}
{Kirk}, J.~G., {Ball}, L., \& {Skj{\ae}raasen}, O. 1999, Astroparticle Physics,
  10, 31. \eprint{astro-ph/9808112}

\bibitem[{{Kishishita} et~al.(2009){Kishishita}, {Tanaka}, {Uchiyama}, \&
  {Takahashi}}]{2009ApJ...697L...1K}
{Kishishita}, T., {Tanaka}, T., {Uchiyama}, Y., \& {Takahashi}, T. 2009, \apjl,
  697, L1. \eprint{0904.1064}

\bibitem[{{Kobulnicky} \& {Fryer}(2007)}]{2007ApJ...670..747K}
{Kobulnicky}, H.~A., \& {Fryer}, C.~L. 2007, \apj, 670, 747

\bibitem[{{Lamberts} et~al.(2012){Lamberts}, {Dubus}, {Lesur}, \&
  {Fromang}}]{2012A&A...546A..60L}
{Lamberts}, A., {Dubus}, G., {Lesur}, G., \& {Fromang}, S. 2012, \aap, 546,
  A60. \eprint{1202.2060}

\bibitem[{{Lamberts} et~al.(2011){Lamberts}, {Fromang}, \&
  {Dubus}}]{2011MNRAS.418.2618L}
{Lamberts}, A., {Fromang}, S., \& {Dubus}, G. 2011, \mnras, 418, 2618.
  \eprint{1109.1434}

\bibitem[{{Lamberts} et~al.(2013){Lamberts}, {Fromang}, {Dubus}, \&
  {Teyssier}}]{2013A&A...560A..79L}
{Lamberts}, A., {Fromang}, S., {Dubus}, G., \& {Teyssier}, R. 2013, \aap, 560,
  A79. \eprint{1309.7629}

\bibitem[{{Lee} et~al.(1991){Lee}, {Osaki}, \& {Saio}}]{1991MNRAS.250..432L}
{Lee}, U., {Osaki}, Y., \& {Saio}, H. 1991, \mnras, 250, 432

\bibitem[{{Liu} et~al.(2006){Liu}, {van Paradijs}, \& {van den
  Heuvel}}]{2006A&A...455.1165L}
{Liu}, Q.~Z., {van Paradijs}, J., \& {van den Heuvel}, E.~P.~J. 2006, \aap,
  455, 1165. \eprint{0707.0549}

\bibitem[{{Martin} et~al.(2014){Martin}, {Nixon}, {Armitage}, {Lubow}, \&
  {Price}}]{2014ApJ...790L..34M}
{Martin}, R.~G., {Nixon}, C., {Armitage}, P.~J., {Lubow}, S.~H., \& {Price},
  D.~J. 2014, \apjl, 790, L34. \eprint{1407.5676}

\bibitem[{{Martin} et~al.(2011){Martin}, {Pringle}, {Tout}, \&
  {Lubow}}]{2011MNRAS.416.2827M}
{Martin}, R.~G., {Pringle}, J.~E., {Tout}, C.~A., \& {Lubow}, S.~H. 2011,
  \mnras, 416, 2827. \eprint{1106.2591}

\bibitem[{{Martin} et~al.(2009){Martin}, {Tout}, \&
  {Pringle}}]{2009MNRAS.397.1563M}
{Martin}, R.~G., {Tout}, C.~A., \& {Pringle}, J.~E. 2009, \mnras, 397, 1563.
  \eprint{0905.2362}

\bibitem[{{McSwain} et~al.(2010){McSwain}, {Grundstrom}, {Gies}, \&
  {Ray}}]{2010ApJ...724..379M}
{McSwain}, M.~V., {Grundstrom}, E.~D., {Gies}, D.~R., \& {Ray}, P.~S. 2010,
  \apj, 724, 379. \eprint{1009.2173}

\bibitem[{{Melatos} et~al.(1995){Melatos}, {Johnston}, \&
  {Melrose}}]{1995MNRAS.275..381M}
{Melatos}, A., {Johnston}, S., \& {Melrose}, D.~B. 1995, \mnras, 275, 381

\bibitem[{{Mignone} \& {McKinney}(2007)}]{2007MNRAS.378.1118M}
{Mignone}, A., \& {McKinney}, J.~C. 2007, \mnras, 378, 1118. \eprint{0704.1679}

\bibitem[{{Moritani} et~al.(2011){Moritani}, {Nogami}, {Okazaki}, {Imada},
  {Kambe}, {Honda}, {Hashimoto}, \& {Ichikawa}}]{2011PASJ...63L..25M}
{Moritani}, Y., {Nogami}, D., {Okazaki}, A.~T., {Imada}, A., {Kambe}, E.,
  {Honda}, S., {Hashimoto}, O., \& {Ichikawa}, K. 2011, \pasj, 63, L25.
  \eprint{1105.4721}

\bibitem[{{Negueruela}(1998)}]{1998A&A...338..505N}
{Negueruela}, I. 1998, \aap, 338, 505. \eprint{astro-ph/9807158}

\bibitem[{{Okazaki}(1991)}]{1991PASJ...43...75O}
{Okazaki}, A.~T. 1991, \pasj, 43, 75

\bibitem[{{Okazaki}(1997)}]{1997A&A...318..548O}
--- 1997, \aap, 318, 548

\bibitem[{{Okazaki} et~al.(2002){Okazaki}, {Bate}, {Ogilvie}, \&
  {Pringle}}]{2002MNRAS.337..967O}
{Okazaki}, A.~T., {Bate}, M.~R., {Ogilvie}, G.~I., \& {Pringle}, J.~E. 2002,
  \mnras, 337, 967. \eprint{astro-ph/0208288}

\bibitem[{{Okazaki} et~al.(2011){Okazaki}, {Nagataki}, {Naito}, {Kawachi},
  {Hayasaki}, {Owocki}, \& {Takata}}]{2011PASJ...63..893O}
{Okazaki}, A.~T., {Nagataki}, S., {Naito}, T., {Kawachi}, A., {Hayasaki}, K.,
  {Owocki}, S.~P., \& {Takata}, J. 2011, \pasj, 63, 893. \eprint{1105.1481}

\bibitem[{{Okazaki} \& {Negueruela}(2001)}]{2001A&A...377..161O}
{Okazaki}, A.~T., \& {Negueruela}, I. 2001, \aap, 377, 161.
  \eprint{astro-ph/0108037}

\bibitem[{{Papaloizou} et~al.(1992){Papaloizou}, {Savonije}, \&
  {Henrichs}}]{1992A&A...265L..45P}
{Papaloizou}, J.~C., {Savonije}, G.~J., \& {Henrichs}, H.~F. 1992, \aap, 265,
  L45

\bibitem[{{Pfahl} et~al.(2002){Pfahl}, {Rappaport}, {Podsiadlowski}, \&
  {Spruit}}]{2002ApJ...574..364P}
{Pfahl}, E., {Rappaport}, S., {Podsiadlowski}, P., \& {Spruit}, H. 2002, \apj,
  574, 364. \eprint{astro-ph/0109521}

\bibitem[{{Pittard}(2009)}]{2009MNRAS.396.1743P}
{Pittard}, J.~M. 2009, \mnras, 396, 1743. \eprint{0904.0164}

\bibitem[{{Porter} \& {Rivinius}(2003)}]{2003PASP..115.1153P}
{Porter}, J.~M., \& {Rivinius}, T. 2003, \pasp, 115, 1153

\bibitem[{{Rajoelimanana} et~al.(2011){Rajoelimanana}, {Charles}, \&
  {Udalski}}]{2011MNRAS.413.1600R}
{Rajoelimanana}, A.~F., {Charles}, P.~A., \& {Udalski}, A. 2011, \mnras, 413,
  1600. \eprint{1012.4610}

\bibitem[{{Reig}(2011)}]{2011Ap&SS.332....1R}
{Reig}, P. 2011, \apss, 332, 1. \eprint{1101.5036}

\bibitem[{{Rivinius} et~al.(2013){Rivinius}, {Carciofi}, \&
  {Martayan}}]{2013A&ARv..21...69R}
{Rivinius}, T., {Carciofi}, A.~C., \& {Martayan}, C. 2013, \aapr, 21, 69.
  \eprint{1310.3962}

\bibitem[{{Sarty} et~al.(2011){Sarty}, {Szalai}, {Kiss}, {Matthews}, {Wu},
  {Kuschnig}, {Guenther}, {Moffat}, {Rucinski}, {Sasselov}, {Weiss}, {Huziak},
  {Johnston}, {Phillips}, \& {Ashley}}]{2011MNRAS.411.1293S}
{Sarty}, G.~E., {Szalai}, T., {Kiss}, L.~L., {Matthews}, J.~M., {Wu}, K.,
  {Kuschnig}, R., {Guenther}, D.~B., {Moffat}, A.~F.~J., {Rucinski}, S.~M.,
  {Sasselov}, D., {Weiss}, W.~W., {Huziak}, R., {Johnston}, H.~M., {Phillips},
  A., \& {Ashley}, M.~C.~B. 2011, \mnras, 411, 1293. \eprint{1009.5150}

\bibitem[{{Sierpowska-Bartosik} \& {Torres}(2009)}]{2009ApJ...693.1462S}
{Sierpowska-Bartosik}, A., \& {Torres}, D.~F. 2009, \apj, 693, 1462.
  \eprint{0811.2466}

\bibitem[{{Stella} et~al.(1986){Stella}, {White}, \&
  {Rosner}}]{1986ApJ...308..669S}
{Stella}, L., {White}, N.~E., \& {Rosner}, R. 1986, \apj, 308, 669

\bibitem[{{Sushch} \& {B{\"o}ttcher}(2014)}]{2014JHEAp...3...18S}
{Sushch}, I., \& {B{\"o}ttcher}, M. 2014, Journal of High Energy Astrophysics,
  3, 18. \eprint{1408.6426}

\bibitem[{{Takahashi} et~al.(2009){Takahashi}, {Kishishita}, {Uchiyama},
  {Tanaka}, {Yamaoka}, {Khangulyan}, {Aharonian}, {Bosch-Ramon}, \&
  {Hinton}}]{2009ApJ...697..592T}
{Takahashi}, T., {Kishishita}, T., {Uchiyama}, Y., {Tanaka}, T., {Yamaoka}, K.,
  {Khangulyan}, D., {Aharonian}, F.~A., {Bosch-Ramon}, V., \& {Hinton}, J.~A.
  2009, \apj, 697, 592. \eprint{0812.3358}

\bibitem[{{Takata} et~al.(2012){Takata}, {Okazaki}, {Nagataki}, {Naito},
  {Kawachi}, {Lee}, {Mori}, {Hayasaki}, {Yamaguchi}, \&
  {Owocki}}]{2012ApJ...750...70T}
{Takata}, J., {Okazaki}, A.~T., {Nagataki}, S., {Naito}, T., {Kawachi}, A.,
  {Lee}, S.-H., {Mori}, M., {Hayasaki}, K., {Yamaguchi}, M.~S., \& {Owocki},
  S.~P. 2012, \apj, 750, 70. \eprint{1203.2179}

\bibitem[{{Tam} et~al.(2014){Tam}, {Kong}, \& {Leung}}]{2014ATel.6216....1T}
{Tam}, P.~H.~T., {Kong}, A.~K.~H., \& {Leung}, G.~C.~K. 2014, The Astronomer's
  Telegram, 6216, 1

\bibitem[{{Tavani} \& {Arons}(1997)}]{1997ApJ...477..439T}
{Tavani}, M., \& {Arons}, J. 1997, \apj, 477, 439. \eprint{astro-ph/9609086}

\bibitem[{{Tavani} et~al.(1994){Tavani}, {Arons}, \&
  {Kaspi}}]{1994ApJ...433L..37T}
{Tavani}, M., {Arons}, J., \& {Kaspi}, V.~M. 1994, \apjl, 433, L37

\bibitem[{{Telting} et~al.(1994){Telting}, {Heemskerk}, {Henrichs}, \&
  {Savonije}}]{1994A&A...288..558T}
{Telting}, J.~H., {Heemskerk}, M.~H.~M., {Henrichs}, H.~F., \& {Savonije},
  G.~J. 1994, \aap, 288, 558

\bibitem[{{The Fermi-LAT Collaboration}(2013)}]{2013arXiv1307.6384T}
{The Fermi-LAT Collaboration} 2013, ArXiv e-prints. \eprint{1307.6384}

\bibitem[{{van Soelen} \& {Meintjes}(2011)}]{2011MNRAS.412.1721V}
{van Soelen}, B., \& {Meintjes}, P.~J. 2011, \mnras, 412, 1721.
  \eprint{1011.2004}

\bibitem[{{van Soelen} et~al.(2012){van Soelen}, {Meintjes}, {Odendaal}, \&
  {Townsend}}]{2012MNRAS.426.3135V}
{van Soelen}, B., {Meintjes}, P.~J., {Odendaal}, A., \& {Townsend}, L.~J. 2012,
  \mnras, 426, 3135

\bibitem[{{Zabalza} et~al.(2013){Zabalza}, {Bosch-Ramon}, {Aharonian}, \&
  {Khangulyan}}]{2013A&A...551A..17Z}
{Zabalza}, V., {Bosch-Ramon}, V., {Aharonian}, F., \& {Khangulyan}, D. 2013,
  \aap, 551, A17. \eprint{1212.3222}

\bibitem[{{Zabalza} et~al.(2011){Zabalza}, {Paredes}, \&
  {Bosch-Ramon}}]{2011A&A...527A...9Z}
{Zabalza}, V., {Paredes}, J.~M., \& {Bosch-Ramon}, V. 2011, \aap, 527, A9.
  \eprint{1011.4489}

\bibitem[{{Zamanov} et~al.(2001){Zamanov}, {Reig}, {Mart{\'{\i}}}, {Coe},
  {Fabregat}, {Tomov}, \& {Valchev}}]{2001A&A...367..884Z}
{Zamanov}, R.~K., {Reig}, P., {Mart{\'{\i}}}, J., {Coe}, M.~J., {Fabregat}, J.,
  {Tomov}, N.~A., \& {Valchev}, T. 2001, \aap, 367, 884.
  \eprint{astro-ph/0012371}

\bibitem[{{Zdziarski} et~al.(2010){Zdziarski}, {Neronov}, \&
  {Chernyakova}}]{2010MNRAS.403.1873Z}
{Zdziarski}, A.~A., {Neronov}, A., \& {Chernyakova}, M. 2010, \mnras, 403,
  1873. \eprint{0802.1174}

\end{thebibliography}

\end{document}